# Hybrid Organic-Metal Oxide Multilayer Channel Transistors with Record Operational Stability

Yen-Hung Lin[1,2]†*, Wen Li[1,3,4]†, Hendrik Faber[1,5], Nikolaos A. Hastas[1,7], Dongyoon Khim[1], Qiang Zhang[5], Xixiang Zhang[5], Nikolaos Pliatsikas[6], Leonidas Tsetseris[7], Panos A. Patsalas[6], Donal D. C. Bradley[2], Wei Huang[3,4], Thomas D. Anthopoulos[1,5]*

[1]Department of Physics and Centre for Plastic Electronics, Blackett Laboratory, Imperial College London, London SW7 2AZ, UK.

[2]Clarendon Laboratory, Department of Physics, University of Oxford, Parks Road, Oxford OX1 3PU, UK.

[3]Institute of Flexible Electronics (IFE), Northwestern Polytechnical University, Xi'an, China

[4]Institute of Advanced Materials (IAM), Nanjing University of Posts and Telecommunications, Nanjing, China.

[5]King Abdullah University of Science and Technology (KAUST), KAUST Solar Centre, Thuwal 23955-6900, Saudi Arabia.

[6]Department of Physics, Aristotle University of Thessaloniki, GR-54124 Thessaloniki, Greece.

[7]Department of Physics, National Technical University of Athens, Athens GR-15780, Greece.

*Correspondence: thomas.anthopoulos@kaust.edu.sa, yen-hung.lin@physics.ox.ac.uk

†These authors contributed equally to this work.

**Abstract**

Metal oxide thin-film transistors are fast becoming a ubiquitous technology for application in driving backplanes of organic light-emitting diode displays. Currently all commercial products rely on metal oxides processed via physical vapor deposition methods. Transition to simpler, higher throughput manufacturing methods such as solution-based processes, are currently been explored as cost-effective alternatives. However, developing printable oxide transistors with high carrier mobility and bias-stable operation has proved challenging. Here we show that hybrid multilayer channels composed of alternating ultra-thin layers (≤4 nm) of indium oxide, zinc oxide nanoparticles, ozone-treated polystyrene and a compact zinc oxide layer, all solution-processed in ambient atmosphere, can be used to create TFTs with remarkably high electron mobility (50 cm$^2$/Vs) and record operational stability. Insertion of the ozone-treated polystyrene interlayer is shown to reduce the concentration of electron traps at the metal oxide surfaces and heterointerfaces. The resulting transistors exhibit dramatically enhanced bias stability over 24 h continuous operation and while subjected to large electric-field flux density (2.1×10$^{-6}$ C/cm$^2$) with no adverse effects on the electron mobility. Density functional theory calculations identify the origin of this enhanced stability as the passivation of the oxygen vacancy-related gap states due to interaction between ozonolyzed styrene moieties and the oxides. Our results sets new design guidelines for bias-stress resilient metal oxide transistors.

**Main text**

Moving away from sophisticated, capital intensive manufacturing processes, soluble semiconductors[1,2,3] not only promise to deliver devices with unusual physical characteristics and enhanced performance, but also trigger a paradigm shift in manufacturing philosophy by embracing scalable, cost-effective processes such as chemical spray pyrolysis,[4] ink-jet printing,[5] slot-die coating,[6] among others. As consequence the interest in solution-based manufacturing of consumer electronics is rapidly increasing with global tech giants investing heavily in emerging forms of printed electronics.[7] Among a variety of soluble electronic materials, oxide semiconductors offer a breadth of intriguing assets, including high charge carrier mobility,[8] optical transparency,[9] versatile synthesis,[10] low manufacturing cost[11] etc., the combination of which makes them ideal for use in a range of rapidly emerging applications in the field of printed electronics. Among them, thin-film transistor (TFTs) technologies are a priority for solution processable oxides as they promise to amplify the technological impact of their vacuum-grown counterparts[11] by reducing the manufacturing cost. For these reasons, continuous research efforts have been devoted to improving the operating characteristics of printable metal oxide TFTs, culminating in performance that surpass even their vacuum-grown counterparts,[10,12] while fast approaching that of polycrystalline-Si TFTs.[4,13-15] Despite this remarkable progress, however, printed metal oxide TFTs still suffer from poor operational instability; a phenomenon commonly attributed to charge-trapping due to the often non-stoichiometric nature of solution-grown metal oxides.[16-18]

Precursor chemistry is an integral part of the quest to eliminate defects in metal oxide semiconducting layers. Looking beyond metal-salt precursors, which represent the state of the art in processing solution-based metal oxide TFTs,[10,12] suspensions of pre-formed nanomaterials, such as nanocrystals (NCs) or nanowires (NWs), are now emerging as promising candidates due to their long range crystallinity and potential for enhanced charge transport.[19,20] Moreover, pre-synthesized nanomaterials do not rely on high temperature annealing often used in conjunction with conventional metal-salt precursors for the in-situ synthesis of metal oxides, as well as for eliminating organic residues[10] that are known to deteriorate charge transport.[21] Despite the promising potential, however, the use of oxide nanomaterials in transistors faces significant challenges that extend beyond layer crystallinity. For example, currently there are a few obstacles that hinder inter-NC (or inter-NWs) charge transport. The first is the presence of insulating stabilizing ligands that prevent efficient electronic coupling between adjacent NCs (or NWs).[20,22] The second is the low density of the layers formed by these nanomaterials, which in turn limit charge percolation and long-range transport.[23,24] To address these challenges, different approaches have been proposed, including ultraviolet ozone (UV-ozone) treatment of the NC layers to remove the ligands and enhance the inter-particle coupling.[20,22] Suspending the nanomaterials in an organic semiconductor binder has also been exploited.[25] In spite of past efforts, however, the performance of metal oxide NC-based TFTs has remained inferior to devices created via the conventional precursor synthesis.[12]

We have recently described an alternative approach for high performance metal oxide TFTs using low-dimensional (ultra-thin) heterojunction (HJ) channels.[4,26,27] In these HJ devices, the engineered energy mismatch between the two semiconducting oxides induces accumulation of electrons in the vicinity of the heterointerface.[27] Resulting transistors outperform those made from individual oxide layer channels, in terms of electron mobility. This is due to a stark transition from a trap-limited conduction to a percolation limited conduction process that takes place in the HJ devices.[27] Despite the enhanced performance, the operating stability of HJ oxide TFTs has so far remained modest due to the persistent electronic defects present in the individual materials and at the heterointerfaces.[16-18] Since the operation of these oxide HJ TFTs relies on the charge carriers transferred from the low-mobility layer on top of the high-mobility layer underneath,[4,26,27] the presence of electronic defects will deteriorate the device performance and its operational stability. Thus, reducing the electron-trap states within the HJ could potentially enhance the electron mobility and bias stability of the device.

Here, we demonstrate a hybrid multi-layer organic/metal oxide transistor channel composed of solution-processed multi-layers of $In_2O_3$, ZnO-NPs, insulating polymer polystyrene (PS) and compact ZnO. We show that incorporation of a UV-ozone treated PS interlayer within the $In_2O_3$/ZnO-NP heterointerface can dramatically reduce the bias-induced instability during continuous transistor operation under high electric flux densities of $2.1 \times 10^{-6}$ $C/cm^2$ for a bias-stress period of 24 h, whilst maintaining the electron mobility >50 $cm^2$/Vs. Moreover, the hybrid lamellar-like channel design enables incorporation of pre-synthesized Al-doped ZnO-NPs that have positive effects on the TFT's operation, hence providing an extra degree of freedom for the design of state-of-the-art transistors processed from solution phase at low temperatures.

We have previously shown that electron transport in ZnO-NPs TFTs is both a percolation and a trap-limited process dictated primarily by the poor inter-NPs coupling and the presence of surface trap states.[20] Recently, the passivation of these surface states has been achieved with the application of organic materials atop the metal oxide.[28-30] To investigate the potential of this approach, we deposited a PS layer atop the ZnO-NPs. The PS surface was then exposed to UV-ozone to reduce its thickness down to a critical dimension (**Supplementary Figure 1**). **Supplementary Figure 2** shows the evolution of the PS thickness as a function of UV-ozone treatment time (0 to 150 s). For the purpose of this study, we fixed the treatment time to 150 s, which resulted in PS layers 3 to 6 nm-thick.

We evaluated the influence of PS on the electron transport across the ZnO-NPs down to the nm-scale using conductive atomic force microscopy (c-AFM) (**Figure 1a**). The current maps for the as-spun (pristine) ZnO-NPs and UV-ozone treated ZnO-NPs/PS layers are presented in **Figure 1b** and **1c**, respectively. Although both samples exhibit similar current variation between high and low conductivity regions, there are consistent differences in the lateral correlation length and absolute current readings. Noticeably, the ZnO-NPs/PS sample appears consistently more conductive than pristine ZnO-NPs (**Figure 1d** and **1e**). Moreover, the pristine ZnO-NPs layer show large lateral features, the size of which reduces substantially for ZnO-NPs/PS samples while the measured current increases threefold (**Figure 1d** and **1e**) despite the identical experimental conditions. These results suggest that the ozone treated PS layer has no adverse effects on current transport in the vertical direction. Excluding the UV-ozone treatment step altogether, leads to highly insulating hybrid ZnO-NPs/PS bilayers.

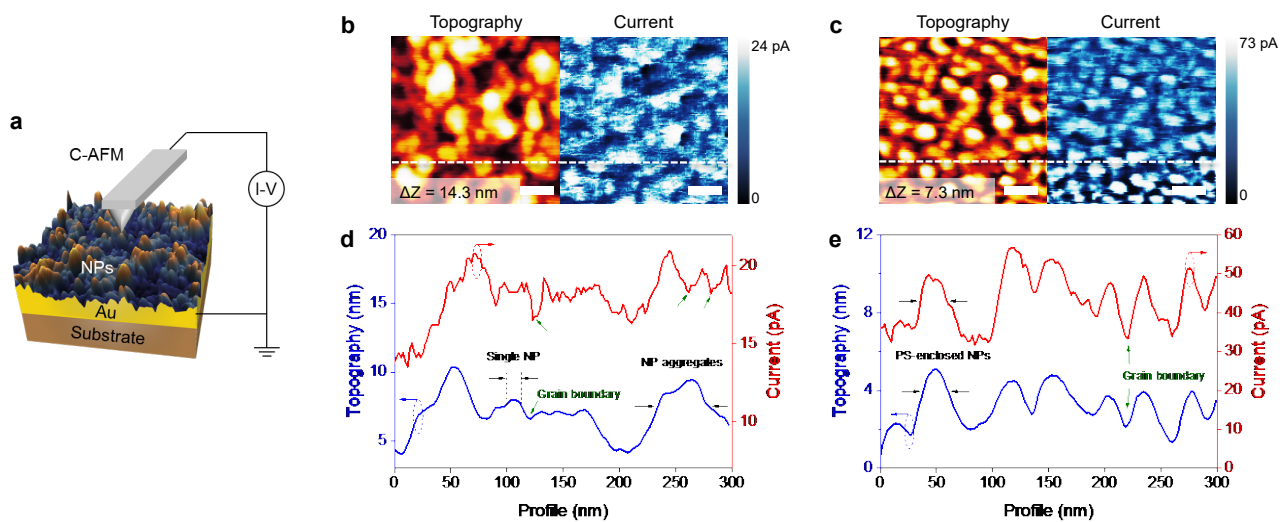

**Figure 1 | Surface topography and current mapping of ZnO NP and NP/PS layers. a**, Schematic of c-AFM measurement setup. **b,c**, Surface topography and current mapping of (**b**) ZnO-NPs and (**c**) ZnO-NPs/PS with surface topography showing a peak-to-peak (z value) amplitude of 14.3 and 7.3 nm, respectively. The scale bar in both images is 50 nm. **d,e**, Cross-sectional profiles [i.e. dashed lines in (**b**) and (**c**)] of the surface topography and current for (**d**) ZnO-NPs and (**e**) ZnO-NPs/PS.

Further insights on the influence of the PS layer come from the intermittent AFM measurements shown in **Supplementary Figure 3**, from which PS appears to 'planarize' the initially rough surface of the ZnO-NPs. This effect is somewhat reversed upon UV-ozone treatment, an observation attributed to gradual etching of the initially thick PS (>10 nm, **Supplementary Figure 2**) and the partial re-exposure of the ZnO-NPs underneath (**Supplementary Figure 3a** and **3b**). Noticeably, many topographic features seen in the ozone treated ZnO-NPs/PS layer are also visible in the c-AFM line scan data shown in **Figure 1e**. We thus conclude that the ozone treated PS layer deposited atop the ZnO-NPs is conformal with an overall positive effect on electron transport (**Figure 1b-1e**). Similar AFM results are obtained for samples of Al-doped ZnO-NPs (ZnO-NPs:Al) as shown in **Supplementary Figure 4**. We note that following the treatment of ZnO-NPs/PS and ZnO-NPs:Al/PS layers with UV-ozone, the smoothening effect (planarization) is manifested as a reduction in the density of visible grains by approximately 50% (**Supplementary Figure 3c**) and 80% (**Supplementary Figure 4c**) for ZnO-NPs/PS and ZnO-NPs:Al/PS, respectively.

To elucidate the origin of the higher current measured by c-AFM for ZnO-NPs/PS layers, we used X-ray photoelectron spectroscopy (XPS) to study the electronic states in the ZnO-NPs and ZnO-NPs/PS layers before and after UV-ozone treatment (**Supplementary Figure 5**). For the pristine ZnO-NPs sample, a minute amount of surface carbon is present. The lineshape of the relevant *C-1s* photoelectron peak indicates that this is exclusively due to adventitious carbon.[31] For the as-spun PS layer, the amplitude of the C-1s peak is substantially enhanced, resembling the typical spectrum of the polymer with a strong contribution at 284.2 eV due to C-C bonds, and a satellite peak close to 291 eV attributed to plasmon losses involving π–π* transitions (red line in **Supplementary Figure 5a**).[32] Following the UV-ozone treatment (blue line in **Supplementary Figure 5a**), the count-rate of *C-1s* is substantially reduced (compared to as-deposited ZnO-NPs/PS (pristine) sample, red line in **Supplementary Figure 5a**), implying a reduced PS layer thickness. Despite this, the amount of carbon in the UV-ozone treated ZnO-NPs/PS sample remains significantly higher than the adventitious carbon observed in as-deposited ZnO-NPs (pristine). We identified the carbonaceous residues following UV-ozone treatment as O−C=O and C−O−C groups (**Supplementary Figure 5a and 5b**),[33] which are consistent with the existence of carbonyls and carboxyls in ozone treated PS, as we will discuss in more detail later. Although we hypothesize that the presence of these chemical species would normally make it difficult to enhance the conductivity of ZnO-NPs, ozone treatment of PS has been shown to affect the operating characteristics of electronic devices.[34,35] In the case of closely packed ZnO-NPs covered with a UV-ozone treated PS layer, the modified surface chemistry could potentially induce various positive effects including trap passivation and enhanced inter-particle coupling.[36]

To study the impact of the UV-ozone treated PS layer on electron transport in the ZnO-NPs layer, we fabricated bottom-gate, top-contact (BG-TC) transistors employing different channel material combinations. **Figure 2a** shows schematics of the various transistor architectures developed while **Figure 2b** displays representative transfer characteristics (*i.e.* drain current, $I_D$, versus gate voltage, $V_G$) for each channel system. **Supplementary Table 1** summarizes the electrical parameters of these transistors. The electron field-effect mobility ($\mu_{SAT}$) calculated for ZnO-NPs and ZnO-NPs:Al transistors, with and without the PS layer on top, are low and in the range of $10^{-3}$-$10^{-4}$ cm$^2$/Vs, in agreement with previous reports.[23,24] Evidently, the presence of the PS layer appears to have little or no effect. In an effort to improve the electronic coupling between the oxide nanoparticles (pristine and Al-doped), a third thin compact layer (≈5 nm) of ZnO was spin-cast atop to form the so-called single nanocomposite channel (SNC) consisting of ZnO-NPs/PS/ZnO(solid) and ZnO-NPs:Al/PS/ZnO(solid) (**Figure 2a**). In both systems the $\mu_{SAT}$ increases to ≈1 cm$^2$/Vs, which is comparable to values obtained for TFTs based on the single layer ZnO(solid) (**Figure 2b**).

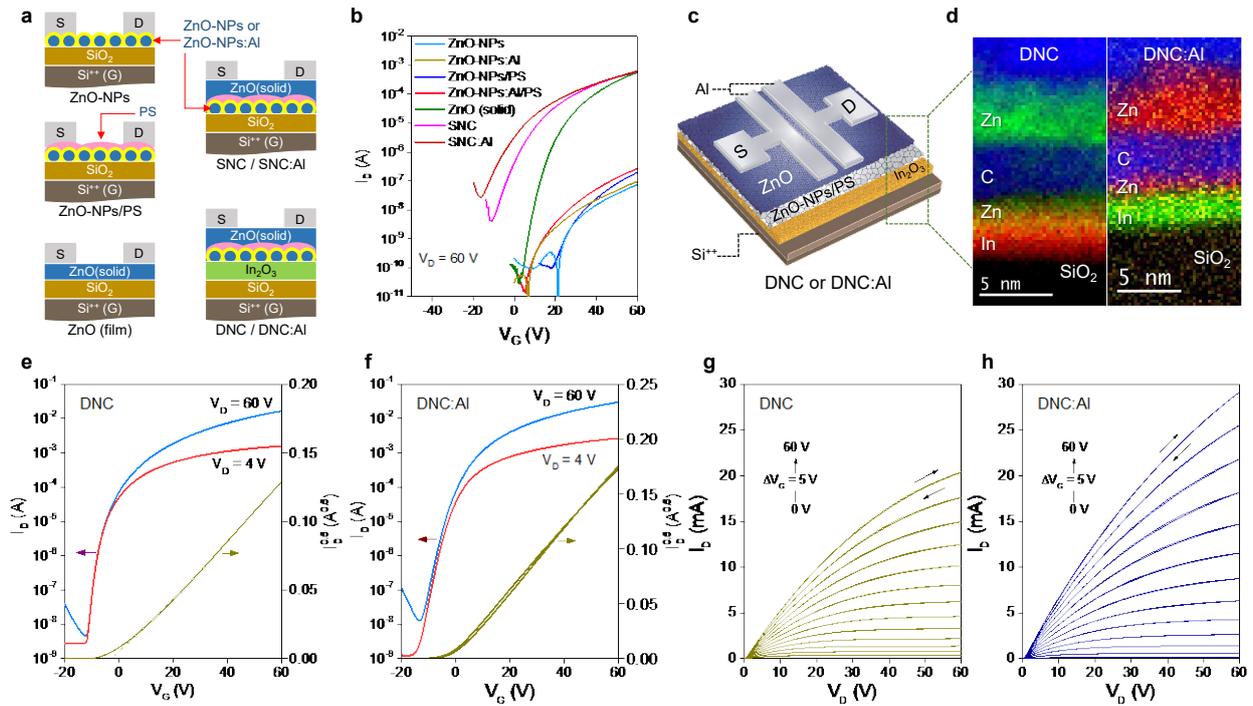

**Figure 2 | Current-voltage characteristics of oxide transistors and EELS analysis of oxide nanocomposites. a**, Schematic of the device structures for transistors based on ZnO-NPs, ZnO-NPs/PS, ZnO and SNC. **b**, Corresponding transfer characteristics for the TFTs shown in (**a**). **c**, Schematic of the DNC device architecture. **d**, Elemental mapping using EELS for the active layers composed of DNC (left) and DNC:Al (right). **e,f,g,h**, Transfer characteristics for (**e**) DNC and (**f**) DNC:Al transistors whilst output characteristics are shown in (**g**) and (**h**), respectively.

The only noticeable difference between the ZnO(solid) and SNC-based TFTs, is the shift in the turn-on voltage ($V_{ON}$) towards more negative $V_G$ seen in the SNC-based devices. This shift is indicative of an increased concentration of free electrons in the channel. We estimated the number of excess electrons by taking the difference in the threshold voltage ($V_{TH}$) between SNC and ZnO(solid) TFTs ($\Delta V_{TH}$) using:

$$\Delta e = \frac{C_i |\Delta V_{TH}|}{q} \quad (1)$$

where, $q$ is the elementary charge and $C_i$ the geometric capacitance of the gate dielectric [20]. Analysis of the data in **Figure 2b** using **Equation 1** yields the areal density of excess electrons of ≈2.71×10$^{11}$ and ≈4.11×10$^{12}$ cm$^{-2}$, for SNC and SNC:Al transistors, respectively (**Supplementary Figure 6**). These results indicate that use of *n*-doped ZnO-NPs:Al induce a higher concentration of electrons in the channel. Measurements of the Fermi energy level ($E_F$) using the Kelvin probe (KP) technique shows that indeed the ZnO-NPs:Al layer has a shallower work function (-3.84 eV) as compared to the pristine ZnO-NPs (-4.23 eV) (**Supplementary Figure 7**), further corroborating the *n*-doped nature of the SNC:Al systems.

Despite the innovative nature of the SNC channel architecture, the transistor performance remains moderate and, in terms of $\mu_{SAT}$, comparable to TFTs based on a conventional single layer ZnO(solid) channel (**Supplementary Table 1**). In an effort to enhance the electron mobility in SNC transistors, we attempted to emulate our recently developed high $\mu_{SAT}$ In$_2$O$_3$/ZnO heterojunction TFTs[4] by replacing the top ZnO layer with the SNC. **Figure 2c** illustrates the schematic of the developed transistor architecture. We hypothesize that in this so-called dual nanocomposite channel (DNC) architecture, the In$_2$O$_3$ layer underneath provides the high electron mobility channel for electrons transferred from the top SNC to the "quantum well" formed in the vicinity of the hybrid In$_2$O$_3$/SNC heterointerface. This is due to the energy band discontinuity between In$_2$O$_3$ and SNC.[4] The accumulated electrons resemble a quasi-2-dimensional electron gas (q2DEG) system.[4,26,27,37]

We first investigated the nature of heterointerfaces formed within the hybrid multilayer channel using scanning transmission electron microscopy (STEM) combined with electron energy loss spectroscopy (EELS). **Figure 2d** shows the elemental maps for the cross-section of the pristine DNC (left panel) and the *n*-doped DNC:Al (right panel) channels, respectively. Both systems appear extremely thin with an overall layer thickness of ≈10 nm. Despite their low dimensionality, the discrete sublayers are clearly visible in both hybrid lamellar structures due to the presence of sharp chemical heterointerfaces. Without any doubt, the EELS data reveal the remarkable potential of solution processing to grow sophisticated hybrid multilayer channels with unprecedented ease and nm-scale precision.

**Figure 2e** and **2f** show representative sets of the transfer current-voltage characteristics measured for DNC and DNC:Al-based TFTs, respectively, whilst the corresponding output characteristics are shown **Figure 2g** and **2h**. For the DNC device, the electron mobility measured in the linear ($\mu_{LIN}$) and saturation ($\mu_{SAT}$) regime are both exceptionally high at ≈36 cm$^2$/Vs and ≈44 cm$^2$/Vs, respectively. For the DNC:Al-based TFT, the mobilities increase even further yielding maximum values for $\mu_{LIN}$ and $\mu_{SAT}$ close to 42 and 51 cm$^2$/Vs, respectively. These values are amongst the highest electron mobilities reported to date for TFTs processed from solution at temperatures ≤200°C.[11,12,27] We attribute this remarkable enhancement to the synergistic effects of trap passivation induced by the ozone treated PS interlayer, and the heterojunction channel architecture employed.

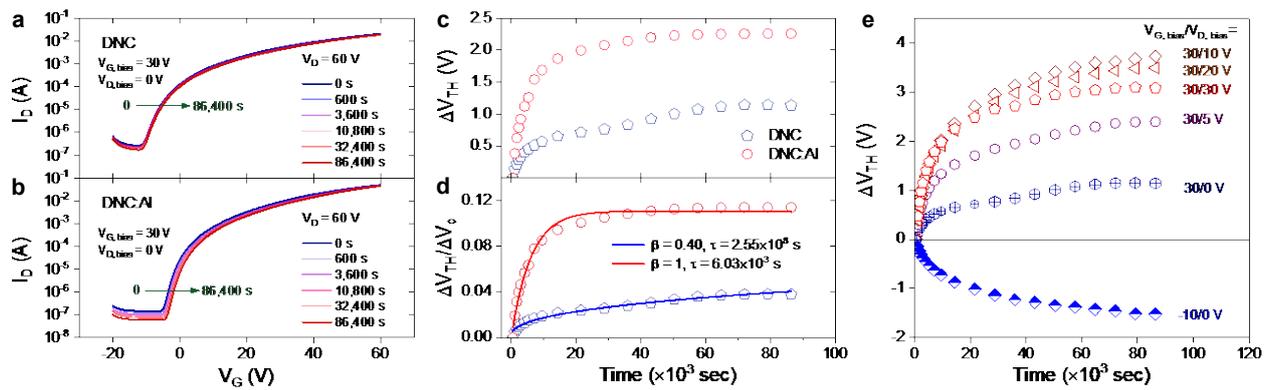

**Figure 3 | Bias-stress measurement and analysis of DNC-based transistors. a,b**, Representative transfer characteristics measured for (**a**) DNC and (**b**) DNC:Al transistors during continuous 24 h bias stress. **c**, Corresponding changes in threshold voltages (ΔV$_{TH}$) for the DNC and DNC:Al devices shown in (**a**). **d**, Experimental data (symbols) and stretched exponential fittings (solid lines) for DNC and DNC:Al transistors. **e**, A series of PBS conditions using a fixed V$_G$ = 30 V and a range of V$_D$ = 0 to 30 V as well as an NBS condition of V$_G$ = -10 V and V$_D$ = 0 V applied to DNC TFTs.

Along with the remarkably high electron mobilities, the DNC transistors are also anticipated to exhibit improved bias stability due to the reduced trap concentration. To investigate the bias stability of DNC-based TFTs, we subjected both types of transistors to continuous positive bias stress (PBS) of V$_G$ = 30 V and V$_D$ = 0 V for 24 h. The representative transfer characteristics for the DNC and DNC:Al devices measured during PBS operation are presented in **Figure 3a** and **3b**, respectively. Remarkably, the channel on-current (I$_{ON}$) deteriorated by 11.7% and 13.1% for the DNC and DNC:Al transistors, respectively, following 24 h of continuous PBS. The associated PBS-induced shift in V$_{TH}$ (ΔV$_{TH}$) is comparatively small and around 1.1 and 2.3 V for the DNC and DNC:Al transistors, respectively. We analysed the PBS results further by fitting the time (*t*)-dependence of $V_{TH}$ with a stretched-exponential equation.

$$\Delta V_{TH}(t) = [V_{TH}(\infty) - V_{TH}(0)] \left[1 - e^{-\left(\frac{t}{\tau}\right)^{\beta}}\right] \quad (2)$$

where, $V_{TH}(0)$ is the threshold voltage prior to bias stress, $V_{TH}(\infty)$ is the threshold voltage when equilibrium has been reached after PBS time ($t \to \infty$) of 24 h, $\tau$ is the characteristic time constant (associated with the charge trapping time), and $\beta$ is the stretching parameter with a numerical value in the range $0 < \beta \leq 1$.[38] A stretching parameter approaching 1 indicates a narrowing distribution of time constants and vice versa when $\beta \ll 1$. Although empirical, **Equation 1** is useful for qualitative comparison of time-dependent charge trapping phenomena that are inherently exponential in their nature but with a distribution of time constants. When a TFT is subjected to continuous bias stress, a value of $\beta \ll 1$ will imply that the time taken for $V_{TH}$ to reach an equilibrium is significantly longer. To this end, $\tau$ represents the time at which $V_{TH}$ has reached 63% of its equilibrium value expected at $t \to \infty$.

In **Figure 3c** we plot the evolution of $\Delta V_{TH}$ as a function of bias-stress time measured at $V_G = 30$ V and $V_D = 0$ V. Evidently, the DNC devices are significantly less sensitive to PBS than DNC:Al transistors, reflected in the noticeably smaller $V_{TH}$ shifts (also seen in **Figure 3a**). In **Figure 3d** we fit the stretched exponential function to $V_{TH}$ as a function of PBS time for the DNC device and obtain $\beta_{DNC} = 0.40$ and $\tau_{DNC} = 2.55 \times 10^8$ s. For the DNC:Al transistor, we find the exponential function (i.e. $\beta = 1$) exhibits significantly better goodness-of-fit than the stretched exponential function and yields a largely reduced $\tau_{DNC:Al} = 6.03 \times 10^3$ s. The apparent increase of the $\beta_{DNC:Al}$ value from the exponential function is in qualitative agreement with the shallower $E_F$ of ZnO-NPs:Al used to construct the DNC:Al channels (**Supplementary Figure 7**). Since electron transport in the latter system is heavily influenced by the dopant-induced states located energetically closer to the conduction band (CB) minimum, the distribution of trapping time constants is expected to narrow, in agreement with the observed increase of the numerical value for $\beta_{DNC:Al}$ (**Figure 3d**).

In **Figure 3e** we show the evolution of $\Delta V_{TH}$ for both types of DNC transistors under PBS at a fixed $V_G = 30$ V and at varying $V_D$ between 0 to 30 V, as well as for a negative bias stress (NBS) at $V_G = -10$ V, $V_D = 0$ V. The corresponding transistor transfer characteristics measured during a bias-stress experiment are shown in **Supplementary Figure 8**. **Supplementary Table 2** summarizes the electrical parameters for DNC transistors characterized before and after 24 h of bias stress at different bias-stress conditions. Noticeably, $\Delta V_{TH}$ decreases with increased $V_D$; a phenomenon attributed to a field-assisted de-trapping of electrons following a process known as the Poole–Frenkel (PF) effect.[38] In the PF mechanism, the lateral source-drain electric field lowers the barrier height for electrons trapped in localized states in the channel, resulting in their release to delocalized transport states. The PF process explains the experimental observation of the diminished $\Delta V_{TH}$ seen under the saturation regime, *i.e.* at $V_D$ = 30 V (**Figure 3e**). Under NBS ($V_G$ = -10 V, $V_D$ = 0 V), the measured $\Delta V_{TH}$ is also small (-1.5 V), despite the long bias-stress time (24 h).

To gain further insight into the origin of the enhanced bias stability of DNC transistors, we investigated the role of the PS interlayer by fabricating a series of devices based on the following channel architectures: (i) $In_2O_3$/ZnO(solid) heterojunction (HJ), (ii) $In_2O_3$/ZnO-NPs/ZnO(solid) (HJ:ZnO-NPs), and (iii) $In_2O_3$/PS/ZnO(solid) (HJ:PS). The schematic of each device architecture is shown in **Figure 4a**. We carried out PBS measurements for all TFTs at $V_G$ = 30 V and $V_D$ = 0 V for a total duration of 24 h. Representative sets of the transfer characteristics measured during bias stressing are shown in **Figure 4b** (HJ), **4c** ($In_2O_3$/ZnO-NPs/ZnO(solid)) and **4d** (HJ:PS), respectively. The corresponding electrical parameters for transistors in different channel configurations measured before and after 24 h of bias stress are summarized in **Supplementary Table 3**. The evolution of the $\Delta V_{TH}$ for these three transistor channel configurations is shown in **Figure 4e,** while the change in $V_{TH}$ (i.e. $\Delta V_{TH}$) for a DNC transistor is also plotted for comparison. Our results reveal the levels of charge trapping between the different channel configurations. The HJ transistor appears to suffer from the most severe bias-induced degradation, as evidenced by the largest $\Delta V_{TH}$. Incorporation of the UV-ozone treated PS interlayer between $In_2O_3$ and ZnO(solid) reduces the bias-stress effect as the $I_{ON}$ exhibited a deterioration of 26.6% compared to 47.1% for the HJ device, hence providing further evidence for the trap passivating role of PS. Meanwhile, the operational stability also sees improvement ($I_{ON}$ deteriorated by 27.2%) when ZnO-NPs are inserted in-between the $In_2O_3$/ZnO(solid) heterointerface to form the HJ:ZnO-NPs transistor. These results confirm that the bias stability of $In_2O_3$/ZnO(solid) heterojunction transistors can be significantly improved by incorporating the ozone treated PS or a ZnO-NPs layer at the vicinity of the heterointerface. Combining the PS and ZnO-NPs interlayers to form the DNC channel system (**Figure 3c**) appears to exhibit an accumulative positive effect, yielding transistors with remarkable bias-stable operating characteristics.

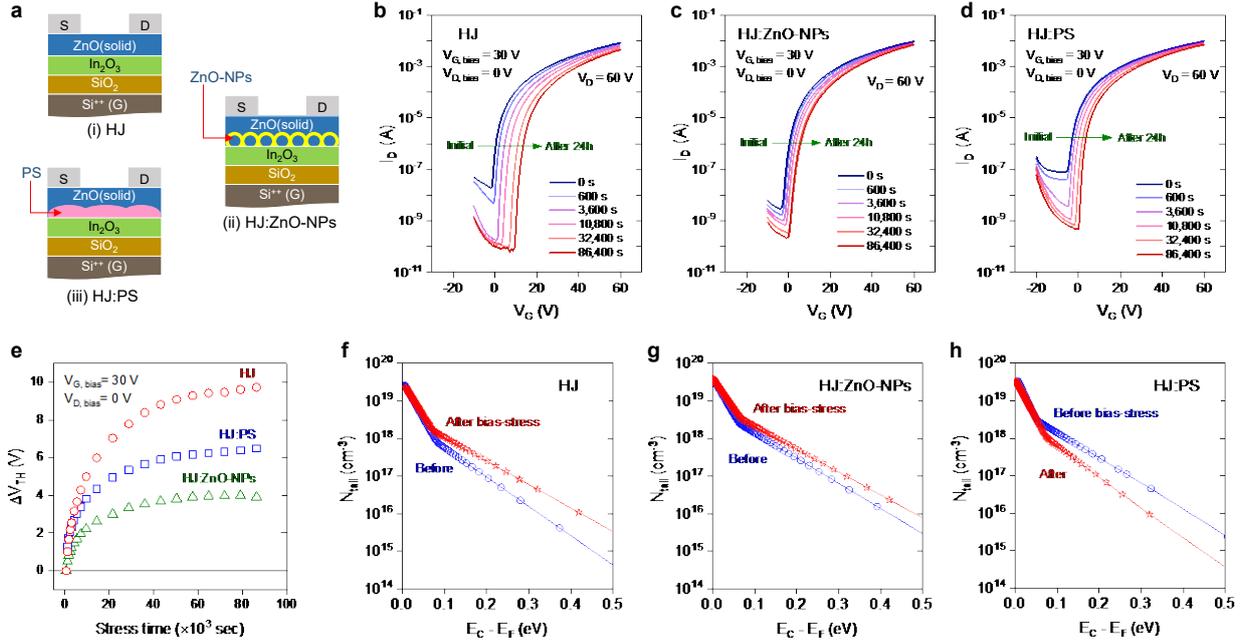

**Figure 4 | Flat-band voltage and trap-state distribution for transistors before/after bias stress. a**, Schematics of the HJ, HJ:ZnO-NPs and HJ:PS transistor configurations. **b,c,d**, Transfer characteristics measured during a 24-h bias-stress condition ($V_G = 30$ V and $V_D = 0$ V) for transistors based on (**b**) HJ (**c**) HJ:ZnO-NPs (**d**) HJ:PS. **e**, Corresponding changes in $V_{TH}$ (i.e. $\Delta V_{TH}$) for the transistors shown in B-D. **f,g,h**, Evolution of density of tail states for (**f**) HJ (**g**) HJ:ZnO-NPs (**h**) HJ:PS before and after bias stress.

To shed more light on the effect of adding ozone treated PS and ZnO-NPs into the HJ transistor channel, the tail state distribution from the CB edge before and after bias stress were calculated using a modified model described by Bubel and Chabinyc.[39-41] The obtained data are presented in **Figure 4f-4h**. Evidently after bias stress, more tail states have been generated in most of the channel systems, similar to conventional single layer metal oxide semiconductor TFTs.[42] While the distributions of the shallow tail states for all the channel configurations remain similar after PBS, for the HJ (**Figure 4f**) and HJ:ZnO-NPs (**Figure 4g**) devices, the concentrations of deep tail states increase upon PBS. Interestingly, the incorporation of ZnO-NPs significantly reduced the generation of the additional tail states induced by bias stress. This result hints that the trapping sites could physically form in-between $In_2O_3$ and ZnO in the HJ, and the ZnO-NPs layer could effectively mitigate this detrimental process. This hypothesis is in good agreement with the smaller shift in $V_{TH}$ observed in **Figure 4e** for the HJ:ZnO-NPs transistor (**Figure 4c**) as compared to a much larger change seen for the HJ device (**Figure 4b**). On the other hand, for the HJ:PS transistor (**Figure 4h**), the addition of the PS layer in-between In$_2$O$_3$ and ZnO seems to reduce the density of deep tail states after PBS. These results, along with the large improvement observed in the reduction of bias stress induced I$_{ON}$ deterioration (**Figure 4d**), confirms the advantageous effect of ozone treated PS layer as a passivation agent in the HJ. Interestingly, similar passivation effects by PS have been reported recently in the field of hybrid perovskite solar cells in order to reduce nonradiative recombination at the perovskite/charge transport layer interface.[43,44] Compared to previous reports on the use of oxide multilayer channels for enhancing the TFT performance,[4,26,27] the hybrid nanocomposite channel architectures demonstrated here provide added advantages the most important of which the exceptional bias stability.

To gain insight into the atomic-scale configurations and mechanisms responsible for the key experimental observations discussed previously, we performed Density Functional Theory (DFT) calculations with the code Quantum Espresso,[45] projector-augmented waves,[46] and a generalized-gradient approximation exchange-correlation (xc) functional.[47] The energy cutoff for the plane-wave basis was set at 75 Rydbergs. We modelled the surfaces of wurtzite ZnO and bixbyite In$_2$O$_3$ with large supercells (typically >120 atoms) as thick slabs with at least 4 layers. The large size of the supercells precluded the possibility of using post-DFT corrections in the form of, for example, a hybrid xc-functional. Sampling of reciprocal space for density of states (DOS) calculations employed the tetrahedron method[48] and 4×4 k-point grids for the surface slabs.

We first examined possible reactions of an ozone (O$_3$) molecule with polystyrene (modelled as a trimer). A typical intermediate configuration when the O$_3$ species is attached to PS is shown in **Supplementary Figure 9a**, whereas **Supplementary Figure 9b** shows the final product of this strongly exothermic reaction with formyl (-CHO) and carboxyl (-COOH) groups. This breakup of an aromatic ring is consistent with experimental reports on UV-ozone treated PS.[49,50] The presence of the -CHO and carbonyl moieties enhances the reactivity of the treated PS significantly. Indeed, as shown in **Supplementary Figure 10**, the treated polymer (modelled without the aromatic rings in order to expedite the calculations without missing key aspects of the structure) chemisorbs on the (0001) surface of ZnO and on the (111) surface of In$_2$O$_3$ through both its formyl and carboxyl groups.

The tendency of treated PS to react with ZnO is an initial, necessary condition for the experimental observations outlined above. To also account for the enhanced conductivity of ZnO NPs (with PS coating) and the bias-stress stability of the DNC transistors, we had to analyse the effect that treated PS has on the properties of the system. For this purpose, we carried out extensive DFT calculations on the electronic properties of ZnO and In$_2$O$_3$ with various native defects in the bulk of these semiconductors and at their surfaces. It transpired that the most relevant surface was the non-polar $(10\bar{1}0)$ termination of ZnO, which has a relatively small formation energy[51] and is thus expected to be present as large facets in ZnO NPs. Oxygen vacancy ($O_V$) is a typical defect for this surface,[51] which is also a common defect in the bulk of the material. As shown in the DOS plot of **Supplementary Figure 11**, an $O_V$ surface defect introduces a state in the band gap of ZnO. Depending on the position of the Fermi level at the $In_2O_3$/ZnO heterostructure, the $O_V$-related gap surface state can be filled or can remain empty. In the latter case, the surface state may act as an electron trap.

The next step was to probe how ozone treated PS interacts with defects in the ZnO surface. Since the UV-ozone treatment creates formyl and carboxyl groups, and in order to accelerate the calculations (again without missing important structural details), we investigated possible reactions of surface oxygen vacancies with formaldehyde and formic acid molecules. As shown in the associated DOS plots of **Supplementary Figure 11**, both molecules eliminate the $O_V$ gap state and shift the Fermi level of the system inside the conduction band. This is a key finding with regard to the experimental data. First, the enhanced conductivity of ZnO NPs, which are coated with treated PS, is consistent with the Fermi energy moving inside the range of the delocalized conduction states. Second, based on the energy band diagram at the $In_2O_3$/ZnO interface, the extra electrons at the bottom of the ZnO conduction band can transfer down in energy to the conduction band of $In_2O_3$, contributing to the formation of the electron-like gas at the interface. Moreover, during positive bias stress, any ZnO oxygen vacancies, which are empty (and thus positively charged), can move toward the ZnO NP facets, where they can be eliminated by the formyl or carboxyl groups of treated PS. The overall result is an enhanced insensitivity of the device to positive bias stress, a prediction that is in qualitative agreement with the bias-stress measurements (**Figure 3** and **4**).

Lastly, we compared the bias-stress stability of our best-performing DNC transistors with most relevant data from the literature. The greatest challenge in performing such a comparison is the unavoidable reliance on literature data measured under different conditions for various device architectures. Since the vast majority of work to date relies on the analysis of $\Delta V_{TH}$ data, accurate comparison between different studies becomes even more challenging. This is because the experimentally measured $\Delta V_{TH}$ depends on various device parameters, such as the chemistry of the gate dielectric, and gate dielectric constant and thickness, which are often overlooked. For this very reason, we have attempted to provide a more holistic comparison by calculating and plotting the three most relevant parameters namely, the electron trap state density (in $cm^{-2}$) induced during bias stressing, the bias-stress time (in s) and the electric flux density (in C/cm$^2$), in a single three-dimensional graph. The combination of these three parameters provides a more reliable basis on the data acquired from different studies.

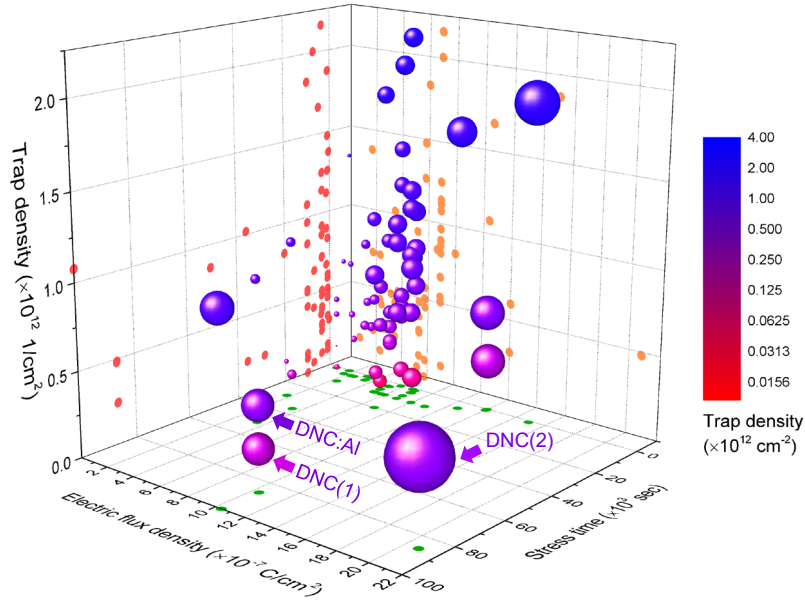

**Figure 5 | 3D scatter plot of transistor bias-stress data.** Data of post bias stress trap density (1/cm$^2$) shown as a function of electric flux density (C/cm$^2$) and stress time (s) for DNC-based transistors and literature data. The size of each dot is proportional to the strength of electric flux density while the projection of each data point on the 2D plane of electric flux density-stress time, trap density-electric flux density and trap density-stress time is shown in green, orange and red, respectively. The bias-stress conditions for the DNC and DNC:Al devices were carried out by applying electric flux densities of 1.04 × 10$^{-6}$ C/cm$^2$ on DNC(1) and DNC:Al, and 2.08 × 10$^{-6}$ C/cm$^2$ on DNC(2) for 24 h. Every bias-stress condition from the literature is summarized in Supplementary **Table 4**.

In **Figure 5** we present the results of our analysis obtained from 60 different sets of bias-stress measurements reported in over 40 different published studies (see **Supplementary Table 4** for more details), against the performance of our DNC-based transistors. To make reading of the data simpler, we made the size of the symbol proportional to the strength of electric flux density, i.e. the larger the dot, the higher the electric flux density applied to the channel. Evidently, both DNC(1,2) and DNC:Al TFTs appear exceptionally resilient towards continuous PBS for up to 86,400 s while being subjected to rather high electric flux densities of 1.04×10$^{-6}$ C/cm$^2$ for DNC(1) and DNC:Al, and 2.08×10$^{-6}$ C/cm$^2$ for DNC(2) transistors. A further intriguing conclusion that can be drawn from **Figure 5** is that most of the published studies rely on bias-stress tests that were undertaken under low electric flux densities ($\leq 1\times 10^{-6}$ C/cm$^2$) and for significantly shorter time periods ($\leq 20\times 10^3$ s). The superior bias stability of the DNC-based TFTs is also reflected in the low trap concentration, which remains constantly below $1\times 10^{12}$ cm$^{-2}$ for our best performing TFTs. This comparison highlights the superior stability of the hybrid organic/metal oxide channel architectures and creates new avenues of research for future generation printable TFT technologies.

In conclusion, we demonstrated solution-processed multilayer channel architectures by combining pseudo-two-dimensional solution processed metal oxides and polystyrene layers with crystalline ZnO nanoparticles. We implemented these layers in TFTs, which produced remarkably high electron mobility (>50 cm$^2$/Vs) and operational stability, even over prolonged periods of continuous operation ($8.64\times 10^4$ s). The key to the success of our channel design is the incorporation of an ozone treated PS interlayer, which passivates the electron traps presented on the surface/interface of metal oxides, and the advantages offered by the heterojunction channel architecture. Use of pre-synthesised Al-doped ZnO nanoparticles has been shown to enable controlled *n*-type doping of the hybrid lamellar channel, providing additional control over the operating characteristics of these high electron mobility TFTs. Detailed investigations of the bias-stress stability, tail-state analysis combined with DFT calculations, have revealed that the proposed hybrid channel design can effectively mitigate the operational instability of pristine oxide channels whilst retaining a high level of device performance. The demonstrated multilayer channel approach could potentially address the numerous shortcomings of single-layer, single-material transistor structures[52] and provide a radically new route towards the development of TFTs with operating characteristics far beyond the current state-of-the-art.

## Method

**Materials and Deposition Techniques.** We purchased ZnO-NPs (Avantama N-10, NP size of 12 nm and concentration of 2.5 wt% in 2-propanol) and ZnO-NPs:Al (Avantama N-21X, 3.15 mol% Al, NP size of 12 nm and concentration of 2.5 wt% in 2-propanol) from Avantama AG. We first diluted the as-received nanoparticles dispersions in 2-propanol to a concentration of 0.2% and then we spin-casted at 3000 rpm for 30 s in ambient air, followed by thermal annealing at 200°C for 10 min. We purchased polystyrene ($M_W = 35,000$) from Sigma-Aldrich and dissolved it in toluene (3 mg/mL) for 24 h at room temperature. We spin-cast the PS solution at 3000 rpm for 30 s in ambient air and thermally annealed at 80°C for 30 min, followed by a UV-ozone treatment step for 150 s. We dissolved indium nitrate [(In(NO$_3$)$_3$] (99.99% Indium Corporation of America) in 2-methoxylenthanol (20 mg/mL) and stirred for 24-48 h at room temperature before deposition. We prepared ZnO by dissolving zinc oxide powder (ZnO, 99%, Sigma-Aldrich) in ammonium hydroxide (50% v/v aq. solution, Alfa Aesar) at a concentration of 8 mg/mL and stirred at room temperature for 24-48 h. We spun the In$_2$O$_3$ and ZnO layers at 6000 and 4000 rpm, respectively, for 30 s in an ambient atmosphere and then thermally annealed them at 200°C for 60 min in air.

**Transistor Fabrication and Characterisation.** We fabricated the transistors in a bottom-gate, top-contact (BG-TC) architecture on heavily-doped Si (Si$^{++}$) wafers with 100-nm thermally-grown SiO$_2$, acting as the gate electrode and gate dielectric, respectively. Prior to thin-film deposition, we thoroughly cleaned the Si$^{++}$/SiO$_2$ wafers using a series of ultra-sonication baths of de-ionized (DI) water with 5% v/v Decon 90 (Decon Laboratories Limited), DI water, acetone and 2-propanol for 10 min each, followed by a UV-ozone treatment step for 10 min. We described the details of the deposition of the TOS thin-film and PS layers in the method section. We completed transistor fabrication with the thermal deposition of a top aluminium (Al) source and drain electrodes under a high vacuum (≈10$^{-6}$ mbar). The channel width (W) and length (L) of the resulting devices were 1000 μm, and 50 μm, respectively. We carried out the electrical current-voltage characterization of the transistors at room temperature in a nitrogen glove box using an Agilent B2902A parameter analyzer. To extract the field-effect mobility from the transfer characteristics, we applied the gradual channel approximation model to the linear and saturation regime using

$$\mu_{LIN} = \frac{L}{C_i W} \cdot \frac{\partial I_D}{\partial V} \cdot \frac{1}{V_D} \qquad (3)$$

$$\mu_{SAT} = \frac{L}{C_i W} \cdot \frac{\partial^2 I_D}{\partial V^2} \qquad (4)$$

where $C_i$ is the geometric capacitance of the gate dielectric used; L and W are the channel length and width, respectively; $V_D$ and $V_G$ are the source-drain voltage and the source-gate voltage, respectively.

**Atomic Force Microscopy.** We measured surface topography information of TOS samples via intermittent contact mode atomic force microscopy (AFM) using an Agilent 5500 AFM system in ambient air. We conducted conducting probe AFM (c-AFM) measurements using the same Agilent 5500 AFM system in contact mode. For the latter measurements, we employed conducting Mikromasch CSC11/Ti-Pt probes with a spring constant in the range of 0.5-6 N/m. We applied a constant tip to sample bias of -3 V during all c-AFM scans.

**Kelvin Probe.** We measured the Fermi energy level of the different NPs using a KP Technology scanning Kelvin Probe system (model number SKP5050) in a nitrogen environment at room temperature.

**Transmission Electron Microscopy.** We acquired high-resolution scanning transmission electron microscopy (HR-STEM) images by an aberration-corrected microscope (Titan G2 60-300, FEI) operated at 300 kV. We characterized the spatial distribution of the elements on the cross-sections of the specimens by electron energy-loss spectroscopy (EELS). We prepared the specimens for STEM characterization by a focus ion beam equipped with a scanning electron microscope dual-beam system (Helios NanoLab 400s, FEI). The final thickness of the specimens was about 50 nm.

**X-ray Photoelectron Spectra.** We acquired core-level (CL) X-ray Photoelectron Spectra (XPS) in a KRATOS Axis Ultra DLD system equipped with a monochromated AlK$_\alpha$ X-ray source, a hemispherical sector electron analyzer and a multichannel electron detector. We acquired the XPS measurements using 20 eV pass energy resulting in a full width at half maximum of the Ag-3d peak less than 500 meV.

**Data Availability:** The data that support the plots within this paper and other finding of this study are available from the corresponding author upon reasonable request.

**Acknowledgement**

Y.-H.L., H.F., D.K., and T.D.A. are grateful to the European Research Council (ERC) AMPRO project no. 280221 for financial support. N.A.H. and T.D.A. are grateful to the European Research Council (ERC) Marie Sklodowska-Curie grant agreement no. 661127 for financial support. The authors are grateful to Kind Abdullah University of Science and Technology (KAUST) for financial support and for facilitating access to the Core Laboratories. L.T. acknowledges support for the computational time granted from GRNET in the National HPC facility – ARIS – under project STEM-2.

**Author contributions**

T.D.A and Y.-H.L. conceived the concept of the project. T.D.A. guided and supervised the project. Y.-H.L. and W.L. fabricated the devices and the thin-film samples and performed the electrical measurements. Y.-H.L. and W.L. analysed all the device data. Y.-H.L. and W.L. carried out the c-AFM, AFM and KP measurements and analysed the data. N.A.H. performed trap state analysis. H.F. and D.K. assisted bias-stress measurement. Q.Z. and X.Z. carried out TEM characterisation. N.P. and P.A.P. carried out XPS characterisation and analysis. L.T. performed DFT analysis. D.D.C.B. and W.H. provided suggestions for material characterisation. Y.-H.L. and T.D.A. wrote the first draft. All authors discussed the results and contributed to the writing of the paper.

## Additional information

Correspondence and request for materials should be addressed to

*thomas.anthopoulos@kaust.edu.sa; yen-hung.lin@physics.ox.ac.uk*

## Competing interests

The authors declare no competing financial interests.